\documentclass[aps,showpacs,prc,twocolumn]{revtex4}
\usepackage{graphicx}
\usepackage{amssymb}
\begin{document}

\title{Dissipative quantum dynamics in low-energy collisions of complex nuclei}

\author{A. Diaz-Torres, D.J. Hinde, M. Dasgupta}
\affiliation{Department of Nuclear Physics, Research School of
Physical Sciences and Engineering, Australian National University,
Canberra, ACT 0200, Australia}
\author{G.J. Milburn}
\affiliation{Centre for Quantum Computer Technology, University of
Queensland, St. Lucia, Queensland 4072, Australia}
\author{J.A. Tostevin}
\affiliation{Department of Physics, Faculty of Engineering and
Physical Sciences, University of Surrey, Guildford, Surrey GU2 7XH,
United Kingdom}

\begin{abstract}
Model calculations that include the effects of irreversible,
environmental couplings on top of a coupled-channels dynamical
description of the collision of two complex nuclei are presented.
The Liouville-von Neumann equation for the time-evolution of the
density matrix of a dissipative system is solved numerically
providing a consistent transition from coherent to decoherent (and
dissipative) dynamics during the collision. Quantum decoherence and
dissipation are clearly manifested in the model calculations. Energy
dissipation, due to the irreversible decay of giant-dipole
vibrational states of the colliding nuclei, is shown to result in a
hindrance of quantum tunneling and fusion.
\end{abstract}
\maketitle

\section{Introduction}

Collisions of composite nuclei involve a complicated interplay and
exchange of energy and angular momentum between the relative motion
and the intrinsic states of the nuclei. High precision data for
low-energy fusion reactions provide one of the most sensitive tests
of such interplay. Stationary state coupled-channels descriptions
have provided a natural methodology to study the effects of specific
excitation modes of one or both of the reactants on the reaction
outcomes. The wave function in this Schr\"odinger coupled-channels
picture is a linear superposition of the states in the model space
with a definite phase relationship. This coherent linear superposition 
can result in enhancement of the quantum tunneling probability and
quantum interference. The coupled-channels approach has been very
successful \cite{Thompson} in explaining several collision
observables. However, problems remain. Foremost is the inability to
describe elastic scattering and fusion measurements simultaneously
\cite{Newton,Angeli} and, related, the more recent failure to 
describe in a physically consistent way the below-barrier quantum 
tunnelling and above-barrier fusion yields \cite{Nanda2}.

The coupled-channels treatment of collisions of complex nuclei as 
closed quantum systems is an approximation. In practice, collisions 
evolve as open quantum
systems, with innumerable bound and continuum intrinsic excitations of the
nuclei. In analogy with problems of a quantum system in a bath, and following 
Bohr and Mottelson \cite{BM75}, we view the nuclear 
many-body Hamiltonian as a sum of collective, single-particle and coupling terms. 
In a nuclear collision, the collective part comprises the relative motion of the nuclei and intrinsic rotational and/or 
vibrational modes. Only a fraction of the Hilbert space of this Hamiltonian is used in {\em any} feasible full coupled-channels calculation. This is because the model space is inevitably restricted to selected, most collective intrinsic excitations. 
These collective states define a {\em reduced} quantum system. 
All other states are {\em weakly} coupled to this reduced system 
by residual interactions and constitute an ``environment". The key question that 
arises is: do environmental effects influence the reaction dynamics and observables, 
such as angular distributions of products or tunneling rates? 

In modelling fusion, the ``environment" is assumed to come into 
play only inside the fusion barrier, and is accounted for in coupled-channels 
calculations through an imaginary potential 
or an ingoing wave bounday condition. However, environmental effects can also be 
manifested before the nuclei encounter the fusion barrier, for example as real or 
virtual excitations of giant resonances in the individual nuclei by the long-range 
Coulomb excitation mechanism. These can be important doorway-states to 
the irreversible loss of kinetic energy (heating of the nuclei), as first 
suggested in Ref. \cite{Broglia} for deep-inelastic reactions. Another mechanism is 
complicated multi-nucleon transfer channels \cite{Rehm}. Measurements have shown 
that deep-inelastic processes occur even at sub-barrier incident energies \cite{Wolfs}, 
in competition with the process of quantum tunneling, and thus fusion \cite{Dasso2}. 
Energy loss associated with the deep-inelastic mechanism thus could play a significant 
role in the inhibition of tunneling at deep sub-barrier energies. 

The investigation of the 
effect, on near- and below barrier fusion, here requires a dynamical model which can include both intrinsic state and environmental couplings in calculations of the tunneling probability. There are no existing realistic theoretical approaches for solving 
this problem. This paper discusses, within a model context, (1) how environment-induced irreversibility can be incorporated into the successful coupled-channels framework, and (2)
makes a first assessment of its effect on a low-energy nuclear collision. The application considered is quantum tunneling, relevant to the low-energy nuclear fusion hindrance phenomenon.

The paper is organized as follows. In Section II we give a brief survey of theoretical approaches to dissipative dynamics of low-energy nuclear collisions, and discuss the suitability of the Lindblad axiomatic theory for the treatment of energy dissipation on sub-barrier fusion. In Section III we present the coupled-channels density matrix approach. Numerical results for our model test case are discussed in Section IV. Finally, the summary of the paper is given in Section V.

\section{Theoretical background}

Neither existing models of fusion nor of deep-inelastic scattering can address both energy 
dissipation and quantum tunneling. The impact of finite lifetimes of excited 
states (e.g., giant resonances) on fusion has been studied within a coupled-channels model 
\cite{Hussein1,Hussein2,HTkawa}, but this approach does not lead to energy dissipation. Direct damped collisions between complex 
nuclei have also been intensively investigated within various approaches, including: 
(i) transport theories \cite{Schroeder} based on premaster, master, Fokker-Planck, 
Langevin and diffusion equations, and (ii) quantum mechanical collective theories 
\cite{Maruhn}. An appealing semi-classical coupled-channels approach combined with a random-matrix model was suggested by Ko \cite{Ko}, which unifies the statistical and coherent pictures of energy dissipation in deep-inelastic collisions. This framework has been succesfully applied \cite{Hussein3, Hussein4} to study the excitation of multiphonon giant resonances in heavy-ion collisions at intermediate energy. In most of these developments the relative motion of the nuclei is described with classical trajectories, whilst the coupling to intrinsic degrees of freedom is treated either statistically (random-matrix theory) or through phenomenological transport coefficients. However, the quantum mechanical treatment of the relative motion is essential for dealing with quantum tunneling.

The quantum dynamical model presented in this work is based on the time evolution
of a reduced density-matrix. It provides a consistent description 
of the transition from a pure state to a mixed quantum state during the
collision. The fundamental equation of motion is the Liouville-von
Neumann equation for an open quantum system, in which a dissipative 
Liouvillian accounts for irreversibility due to interactions 
of the system with an environment. The Lindblad 
axiomatic approach \cite{Lindblad,Sandulescu} for open quantum systems 
has been successfully applied in nuclear physics, but within rather schematic models. 
For instance, to describe the charge 
equilibration process in deep-inelastic collisions 
\cite{Scheid1}, fission \cite{Stefanescu1}, decay of giant resonances 
\cite{Stefanescu2}, tunneling through a parabolic barrier 
\cite{Adamian1}, and scattering in a two-dimensional inverse parabolic 
potential \cite{Genkin}. These are calculations for a {\em single} channel of 
either one damped oscillator \cite{Stefanescu1,Stefanescu2,Adamian1} or 
two coupled damped oscillators \cite{Scheid1,Genkin}.   

In low-energy nuclear collisions, the 
context of the present application, Lindblad's dynamics for the 
evolution of the reduced system is justified (i) because the coupling 
to the complex environment (through excited doorway-states and 
determined by residual interactions) is weak, and 
(ii) because the Markov approximation is expected to be valid, the collective motions of
the two nuclei being slower than the rearrangement
of the environmental (nucleonic) degrees of freedom. With increasing collision energy, to well-above the Coulomb barrier, memory effects related to diabatic dynamics \cite{Noerenberg,Alexis0,Alexis1} may be 
important and a non-Markovian Liouvillian may be required. The weak coupling between two 
subspaces of the total space of intrinsic nuclear states distinguishes 
between the system and environment. The Lindblad theory does not require \cite{Dietz} any limitation on the strength of the system-environment coupling, although the definition of physically well-defined environment states would require a careful analysis in the strong-coupling limit.

An essential effect of the environment
on the reaction dynamics (unlike the effect of absorptive terms) is
to progressively destroy the coherent linear superposition and the associated
phase relationships between different channels, introducing quantum 
decoherence in the system. 

Here, we identify two such (model) sources of decoherence 
and dissipation. Firstly, an environment inside the Coulomb barrier, which is 
related to the complexity of compound nucleus states. Secondly, one with effectively 
a long range, associated with decay out of short lived (compared to the reaction time) 
internal vibrational states, e.g. the giant dipole
resonance (GDR) of the colliding nuclei, will be shown to be of
particular importance. The damping of the GDR, due to its irreversible
coupling to a sea of complicated surrounding states, which constitute
the environment \cite{Bertsch}, destroys the coherent dynamical
coupling with the relative motion of the nuclei. Here we show that
damping of the GDR results in decoherence and energy loss in the region where 
the nuclei overlap, inhibiting tunneling, and thus fusion.
      
\section{Coupled-channels density matrix approach}

We exploit the time-evolution of a coupled-channels density matrix,
as is employed in quantum molecular dynamics \cite{Pesce1}. The
density operator $\hat{\rho}$ in Eq.\ (\ref{eq1}) is represented in
an asymptotic (product) basis of states of the internal Hamiltonian of
the individual nuclei, $|i \rangle, i = 1,\ldots N$ (lower indices),
and coordinate states describing the separation of the two nuclei,
$|r), r = 1,\ldots M$ (upper indices). That is,
\begin{equation}
\hat{\rho} = \sum_{ij,rs} |r ) |i \rangle \, \rho_{ij}^{rs} \,
\langle j| ( s| \label{eq1}\ .
\end{equation}
Crucially, we also add two auxiliary states to the $|i \rangle$
basis that allow distinct environmental interactions, as described
below. The density operator obeys the time-dependent Liouville-von
Neumann equation
\begin{equation}
\frac{\partial \hat{\rho}}{\partial t} = \hat{\mathcal L}\hat{\rho}
= [\hat{\mathcal L}_H + \hat{\mathcal L}_D]\hat{\rho} , \label{eq2}
\end{equation}
where the total Liouvillian consists of a Hamiltonian part
$\hat{\mathcal L}_H \hat{\rho}= -{i}[\hat{H}, \hat{\rho}]/{\hbar}$
describing the coherent evolution of the system with Hamiltonian
$\hat{H}$, and a dissipative part $\hat{\mathcal L}_D$ accounting
for the interactions with the environment. Here, $\hat{\mathcal
L}_D$ will be assumed to be given by Lindblad's dissipative
Liouvillian \cite{Lindblad,Sandulescu} associated with a
Markovian semigroup evolution, i.e.
\begin{equation}
\hat{\mathcal L}_D \hat{\rho} = \sum_\alpha \bigl( \hat{\mathcal
C}_\alpha \, \hat{\rho} \, \hat{\mathcal C}_\alpha^{\dag} -
\frac{1}{2} \bigl[\hat{\mathcal C}_\alpha^{\dag} \, \hat{\mathcal
C}_\alpha ,\hat{\rho} \bigl]_{+} \bigl)\ , \label{eq3}
\end{equation}
where $[\ldots]_{+}$ denotes the anti-commutator. Here each
$\hat{\mathcal C}_\alpha$ is a Lindblad operator for a dissipative
coupling, physically motivated according to the specific problem. We
assume that each coupling $\alpha \equiv (Ij)$ between a given state
$|j \rangle$ and an environmental state $|I \rangle$ has an
associated rate $\Gamma_{Ij}$, i.e, $\hat{\mathcal C}_{Ij}
=\sqrt{\Gamma_{Ij}} |I \rangle \langle j|$ \cite{Saalfrank2},
determined by the inverse lifetime of the excited states and the
branching ratio of its de-excitation, taken to be that when the
nuclei are well-separated. We note also that: (a) the Lindblad
Liouvillian has been derived using microscopic models
\cite{Stefanescu2,Liouvillians}, and (b) in contrast to many (dissipative) model
Liouvillians \cite{Sandulescu}, Eq.\ (\ref{eq3}) preserves both the
positivity and the trace of the density matrix. These are essential
properties in any realistic application.

In the model calculations that follow, the basis will comprise two
asymptotic states (coupled-channels) $|1 \rangle$ and $|2 \rangle$
with channel energies $e_j$. Channel $|1 \rangle$ is the (ground
states) entrance channel and is coupled to an inelastic state $|2
\rangle$ by a coupling interaction $V_{12}$. Two distinct sources of
irreversibility are also considered, modelled by two auxiliary
(environment) states $|X \rangle$ and $|Y \rangle$. The first
environmental coupling describes capture by the potential pocket
inside the fusion barrier. This simulates the irreversible and
dissipative excitations associated with the evolution from the two
separate nuclei to a compound nuclear system. In a stationary states
approach this loss of flux is approximated by imposing an imaginary
potential $-iW(r), W(r)>0,$ or an ingoing-wave boundary condition at
distances well inside the barrier. Here, these transitions are
described by an auxiliary state $|X \rangle$, to which \emph{all}
other states $|j \rangle $ couple, modelled \cite{Irene} by a
Lindblad operator $\hat{\mathcal C}_{Xj}=\sqrt{\gamma ^{rr}} |X
\rangle \langle j|$. The absorption rate to state $|X \rangle$ is
given by $\gamma^{rr} = W(r)/\hbar$ where $W(r)$ is taken as a Fermi
function with depth 10 MeV and diffuseness 0.1 fm, located at the
pocket radius of the nucleus-nucleus potential, $\approx 7$ fm. This
choice guarantees complete absorption inside the pocket. The fusion
probability is defined as the probability accumulating in this state
$|X \rangle $.

The second environment, whose explicit treatment will be seen to be
the most significant at lower energies, is associated with the
irreversible decay out of intrinsic excitations of the colliding
nuclei. Such decays are independent of the dynamical couplings.
Specifically, we will associate the only excited coupled channel
state $|2 \rangle$ with the GDR excitation. We then introduce a
second auxiliary state $|Y \rangle$, representing the bath of states
in which the GDR is embedded, and to which only the GDR excitation
$|2 \rangle$ is coupled.

\begin{figure}
\begin{center}
\includegraphics[width=0.40\textwidth,angle=0]{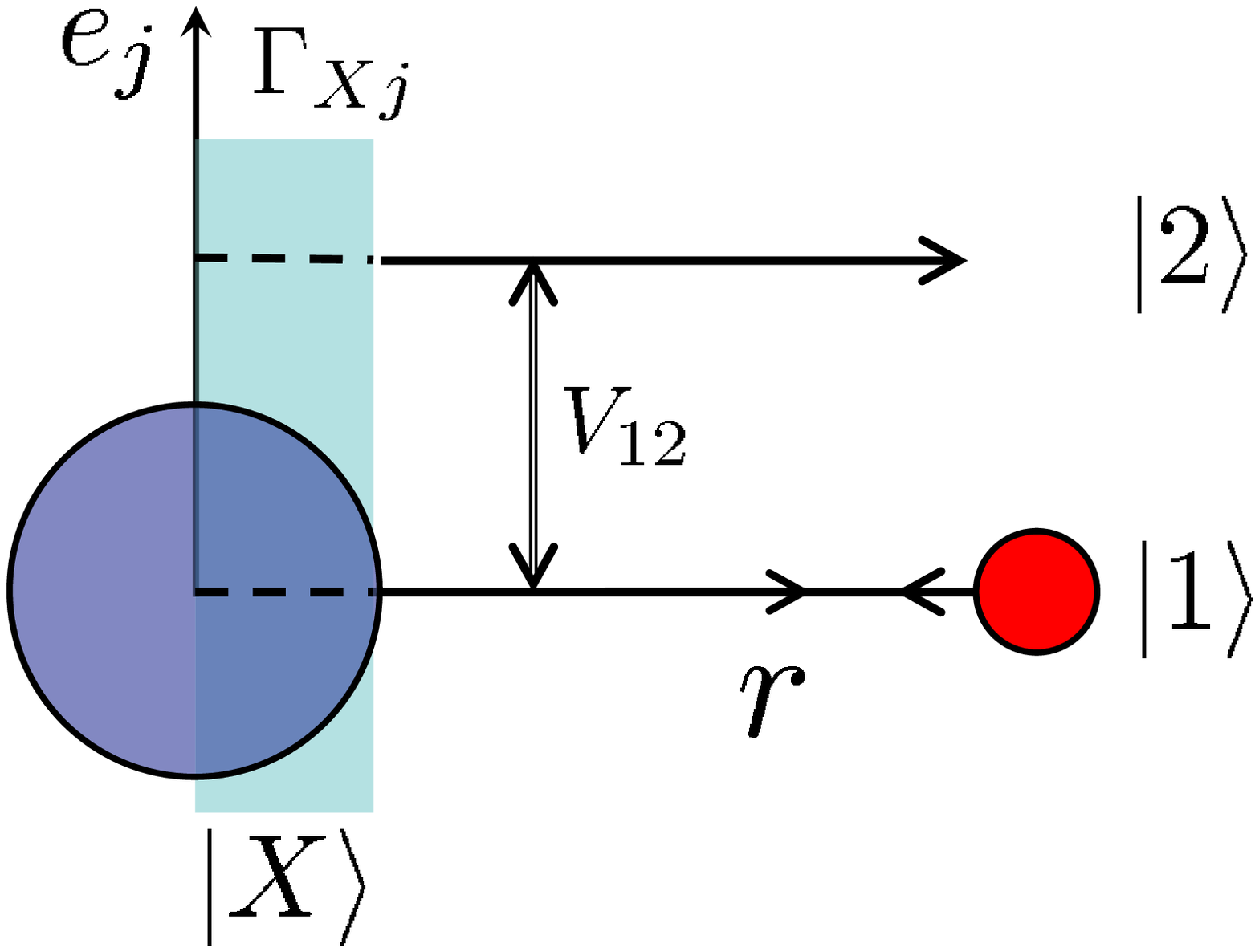} 
\includegraphics[width=0.40\textwidth,angle=0]{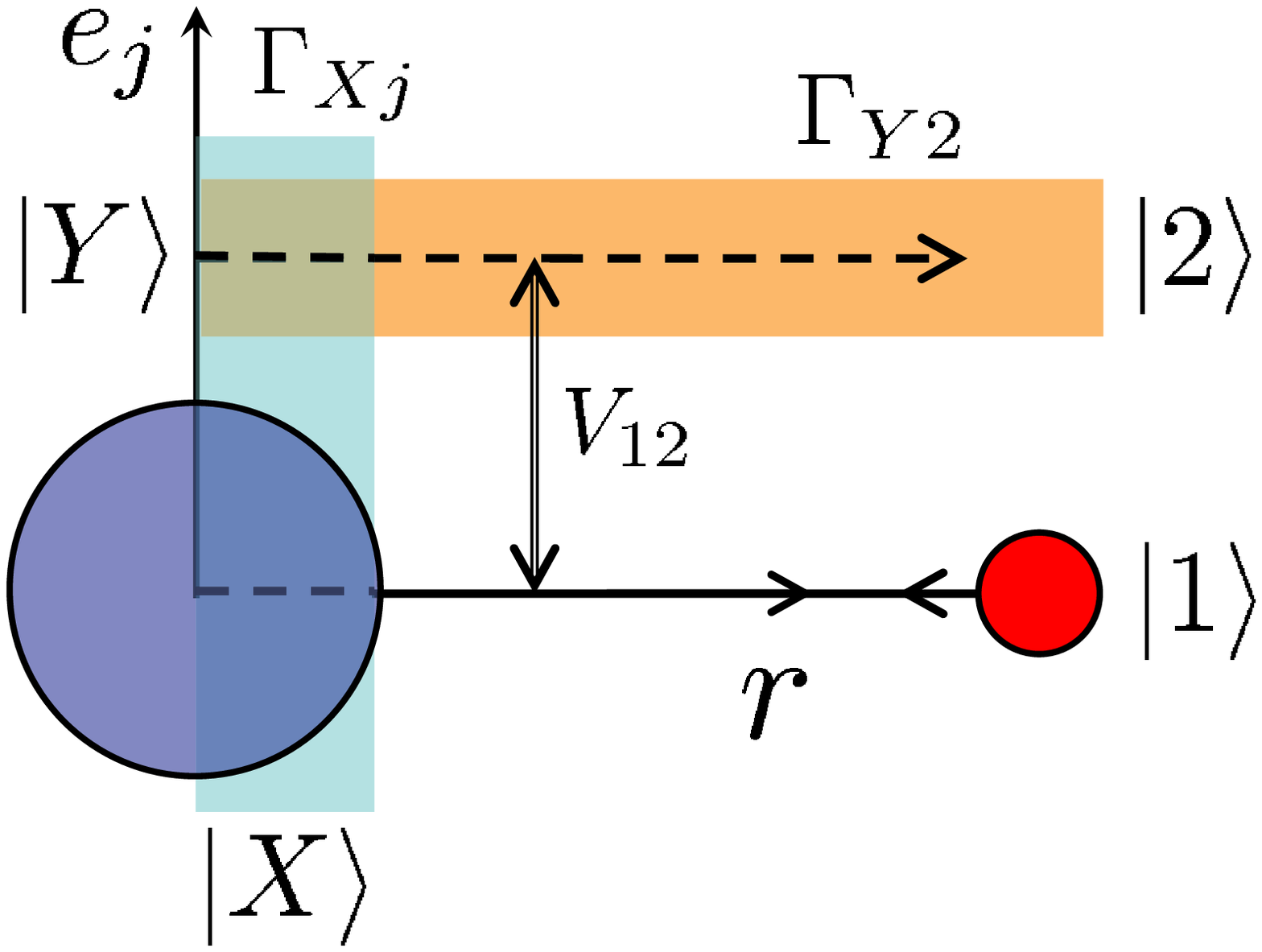}
\caption{(Color online) Schematic representations of the dissipative
coupled channels model calculations, with channel energies $e_j$,
showing the spatial and energy localization of the auxiliary
(environment) states $|X \rangle$ and $|Y \rangle$ (shaded rectangles).
The upper panel shows the dynamical calculation scheme in the
presence of environment $|X \rangle$ only. The lower panel shows the
dynamical calculation scheme in the presence of both environments
$|X \rangle$ and $|Y \rangle$. Here, dashed lines indicate regions
where the intrinsic channels $|1 \rangle$ and $|2 \rangle$
experience irreversible  couplings to the environmental states $|X
\rangle$ and/or $|Y \rangle$.} \label{Figure0}
\end{center}
\end{figure}

Thus, $|Y \rangle$ and/or $|X \rangle$ supplement the two intrinsic
states $|1 \rangle$ and $|2 \rangle$ that comprise the two coupled
channels. Both of the auxiliary states refer to complex excitation
modes of the nuclei, associated with nucleonic degrees of freedom
and compound nucleus states, respectively. They provide intuitive
and formal channels \cite{Irene} for describing irreversible
coupling and loss of probability from the system to these
environments, couplings that enter \emph{only} through the
dissipative dynamics term $\hat{\mathcal L}_D$ in Eq.\ (\ref{eq2}).
$|Y \rangle$ is also assumed to couple to $|X \rangle$ at the
appropriate range of separations. Probability accumulating in state
$|Y \rangle$ outside of this $|X \rangle$ pocket may be identified
with deep-inelastic processes, as will be discussed later. These
environments and the couplings present in the model calculations are
represented schematically in Figure \ref{Figure0}. There, dashed
lines indicate regions where the intrinsic coupled-channels states
$|1 \rangle$ and $|2 \rangle$ experience irreversible (environment)
couplings to states $|X \rangle$ and/or $|Y \rangle$.

Upon inserting Eq.\ (\ref{eq1}) into Eq.\ (\ref{eq2}), the following
coupled equations are obtained for the time evolution of the
density-matrix elements,
\begin{equation}
\dot{\rho}_{ij}^{rs} = (\hat{\mathcal L}_H \hat{\rho})_{ij}^{rs} +
(\hat{\mathcal L}_D \hat{\rho})_{ij}^{rs}\ . \label{eq4}
\end{equation}
Explicitly, the Hamiltonian terms are given by
\begin{eqnarray}
(\hat{\mathcal L}_H \hat{\rho})_{ij}^{rs}&&= -\frac{i}{\hbar}
\biggl[ \rho_{ij}^{rs}(e_i - e_j) + \sum_{t=1}^M
(T^{rt}\rho_{ij}^{ts} -
\rho_{ij}^{rt}T^{ts}) + \nonumber \\
&&\rho_{ij}^{rs}(U^{rr} - U^{ss}) + \sum_{k=1}^N
(V_{ik}^{rr}\rho_{kj}^{rs} - \rho_{ik}^{rs}V_{kj}^{ss}) \biggl]
\label{eq5}
\end{eqnarray}
where $i,\ j$ and $k$ run only over intrinsic states $|1 \rangle$,
$|2 \rangle$, $\ldots$. The dissipative terms are given by
\begin{equation}
(\hat{\mathcal L}_D \hat{\rho})_{ij}^{rs} = \delta_{ij} \sum_{k}
\widetilde{\Gamma}_{ik}^{rr} \rho_{kk}^{rs} -\frac{1}{2} \biggl[
\sum_{k}(\widetilde{\Gamma}_{ki}^{rr} +
\widetilde{\Gamma}_{kj}^{rr}) \biggl] \rho_{ij}^{rs}\ , \label{eq6}
\end{equation}
where the indices run over all the included intrinsic state to
auxiliary state couplings and $\widetilde{ \Gamma}_{ij}^{rr} =
\Gamma_{ij} + \gamma ^{rr}$. In Equation\ (\ref{eq5}), $T$, $U$ and
$V$ refer to the relative kinetic energy, the total bare
nucleus-nucleus potential (Coulomb + nuclear) and the coupling
interaction between the intrinsic channels, respectively.

While not the technique that is used here, we note that this
Lindblad dynamical model can also be recast and solved numerically
within the Monte Carlo wave function method, see e.g. 
Ref. \cite{Molmer}. In that approach, decoherence and 
dissipation originate from the introduction of random quantum jumps 
in the time evolution of the wave function of the system. 
This unraveling density-matrix evolution, through stochastic wave-function 
methods \cite{Molmer}, shows that the two approaches are equivalent and 
the former takes account of the role of fluctuations in the calculation 
of the expectation values and variances of observables.  

The present calculational scheme is based on Eqs.\ (\ref{eq5}) and
(\ref{eq6}) and proceeds as follows. Initially, at time $t=0$, the
nuclei are well-separated, in their ground states, and their density
matrix describes a pure state with Tr$[\hat{\rho}]$ = Tr$[\hat{
\rho}^2] = 1$. An initial wave-packet describes the relative motion
of the nuclei. The coupled equations are solved numerically using
the Faber polynomial expansion of the time-evolution superoperator
\cite{Faber}, $\exp (\tau \hat{\mathcal L})$, and the Fourier method
of Ref. \cite{Kosloff2} for the commutator between the kinetic
energy and density operator. Having solved for the dynamical
evolution of the density matrix, expectation values of an observable
$\hat{\mathcal O}$ are now obtained from the trace relation $\langle
\hat{\mathcal O (t)} \rangle  = \textnormal{Tr}[\hat{\mathcal
O}\hat{\rho}(t)]$. The purity of the initial state, conserved under
Hamiltonian unitary evolution, will be destroyed (Tr[$\hat{\rho}^2]
< 1$) if the environment causes a loss of quantum coherence. This
decoherence can thus be quantified via this loss of density-matrix
purity, or equivalently by an increase of the linear entropy $1-
$Tr$[\hat{ \rho}^2]$.

\section{Numerical results and discussion}

So as to make contact with the coherent, Schr\"odinger picture, the
model Hamiltonian we used was chosen to coincide with that of the
coupled-channels fusion model {\sc ccfull} \cite{Hagino}.
Specifically, our model calculations use physical parameters
relevant to the $^{16}$O + $^{144}$Sm reaction at collision energies
below its nominal fusion barrier, $V_B = 61.1$ MeV. We assume zero
relative orbital angular momentum between the reactants. The form of
the bare nuclear potential between the two nuclei, consistent with
the stated $V_B$, is a Woods-Saxon potential with ($V_0, r_0,a_0)
\equiv$ ($-$105.1 MeV, 1.1 fm, 0.75 fm). The Coulomb potential was
that for two point charges. The $^{16}$O projectile was taken to be
inert and the $^{144}$Sm target was allowed to be excited to a GDR
vibrational state. The dynamical nuclear coupling of the ground
state $|1 \rangle$ to the vibrational state $|2 \rangle$, with
excitation energy $E_{1^{-}}=15$ MeV, has a macroscopic deformed
Woods-Saxon form with a deformation parameter of $\beta_1=0.2$.

The time step for the density-matrix propagation was $\Delta t =
10^{-22}$ s, and the radial grid ($r=0-250$ fm) was evenly spaced
with $M = 512$ points. The relative motion of the two nuclei in the
entrance channel $|1 \rangle$ was described by a minimal-uncertainty
Gaussian wave-packet, with width $\sigma_0 = 20$ fm, initially
centered at $r=150$ fm, and was boosted towards the target with the
appropriate average kinetic energy for the entrance channel energy
$E_0$ required. The FWHM energy spread of the wave-packet is $\sim 3
\%$. The numerical accuracy of the time evolution was checked using
a fully coherent, time-dependent calculation, excluding coupling to
states $|X \rangle $ and $|Y \rangle $. It was confirmed that the
normalisation and purity of the density-matrix, Tr$[\hat{\rho}]$ =
Tr$[\hat{\rho}^2]=1$, and the expectation value of the system energy
Tr$[\hat{H} \hat{\rho}]$ were maintained with high accuracy over the
required number of time steps, typically 700 for the full duration
of the collision.

\begin{figure}
\begin{center}
\includegraphics[width=0.46\textwidth,angle=0]{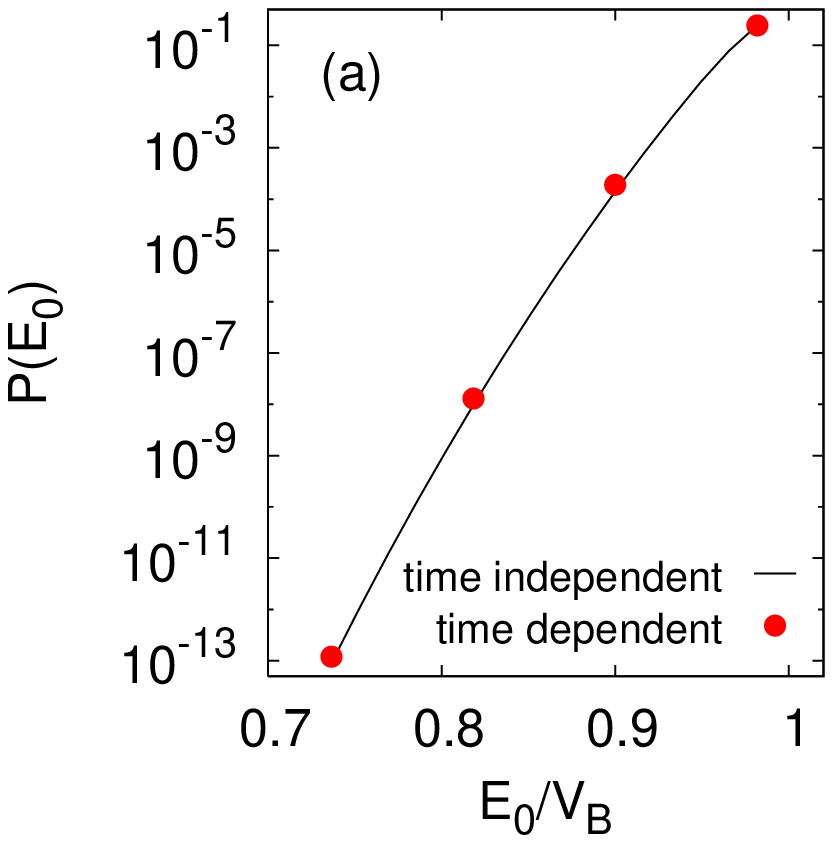} 
\includegraphics[width=0.39\textwidth,angle=0]{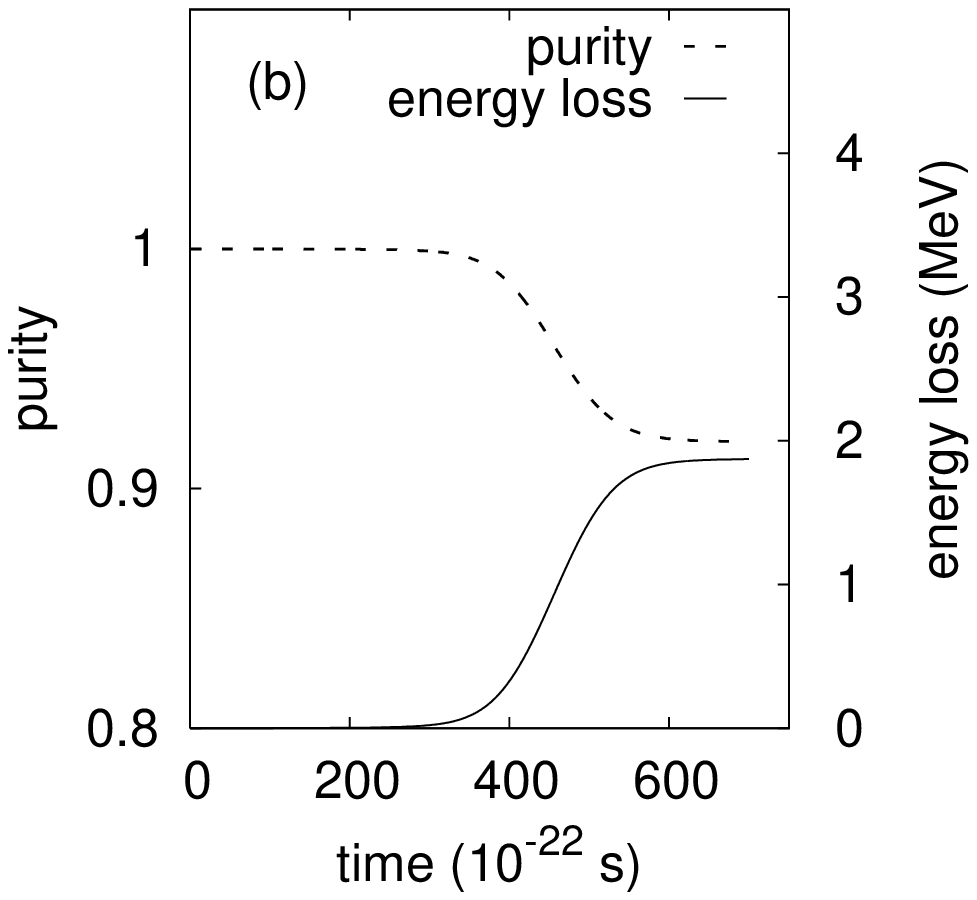}
\caption{(Color online) (a) The energy dependence of the s-wave
tunneling probability calculated with the density matrix (solid
points) and the coupled-channels {\sc ccfull} methods (full line).
(b) The time evolution of the density matrix purity Tr$[\hat{
\rho}^2$] and the energy loss Tr$[\hat{H}(\hat{\rho_0}-\hat{\rho})]$
with decoherent $|X \rangle $ and $|Y \rangle $ state dynamics:
corresponding to the right-hand $E_0=45$ MeV entry in Table 1.}
\label{Figx}
\end{center}
\end{figure}

The importance of the two, spatially distinct, sources of
environment couplings were studied. Calculations were first
performed in the scheme shown in the left panel of Figure
\ref{Figure0}. Here the intrinsic coupled channels $|1 \rangle$ and
$|2 \rangle$ also couple to the capture state $|X \rangle $.
Calculations were carried out for $E_0$ = 45, 50, 55 and 60 MeV
incident energy. The calculated state purity Tr$[\hat{\rho}^2]$ and
the energy dissipation Tr$[\hat{H} (\hat{\rho_0}- \hat{\rho})]$ post
the collision (after 700 time steps) are shown in the left panels in
Table \ref{Table}. For sufficiently sub-barrier energies, $E_0\leq
55$ MeV, it is evident that time-evolution in the presence of state
$|X \rangle$ essentially maintains coherence and is non-dissipative.
There is however loss of purity and dissipation at the highest
energy. It is interesting therefore to compare the density-matrix
and the Schr\"odinger predictions of {\sc ccfull} (that uses an
ingoing wave boundary condition). This is done here only for
calculations of the tunneling probability $P(E_0)$, in a relative
s-wave, shown in Figure \ref{Figx}(a). These comparisons, of
necessity, require convolution of the $\ell=0$ partial wave
penetrabilities $T_0(E)$ from {\sc ccfull} with the energy
distribution $f(E,E_0)$ of the chosen initial wave packet. That is,
$P(E_0) \equiv \int dE\, f(E,E_0)\, T_0(E)$. The $P(E_0)$, shown as
a function of $E_0/V_B$ in Figure \ref{Figx}(a), are in very good
agreement showing the appropriateness of stationary state
coupled-channels calculations for this observable within the
dynamical scheme of states $|1 \rangle$, $|2 \rangle$ and $|X
\rangle$. It is our contention that the dissipation associated with
state $|X \rangle$, while significant at 60 MeV, is strongly
localised inside the barrier and thus does not impact upon the
barrier penetrability. We will now show that the same is not true
for the more spatially-extended dissipation due to the GDR decay
environment $|Y \rangle$.

\begin{table} \caption{The calculated density matrix purity
Tr$[\hat{\rho}^2]$ and energy loss $\Delta E =$Tr$[ \hat{H}
(\hat{\rho_0}- \hat{\rho})]$ following time-evolution (for 700 time
steps) when including only the state $|X \rangle$ (left entries) and
both states $|X \rangle $ and $|Y \rangle$ (right entries)
environmental couplings. The GDR coupling strength used was
$\beta_1=0.2$. } \label{Table}
\medskip
\begin{center}
\begin{tabular}{c|cc|cc}
\hline\hline &\multicolumn{2}{c}{State $|X \rangle $} &
\multicolumn{2}{|c}{States $|X \rangle $ and $|Y\rangle$}
\\
\hline
$E_0$ (MeV) &~~~Tr$[\hat{\rho}^2]$~~&~~$\Delta E$ (MeV)~~ &~~ Tr$[\hat{\rho}^2]$~~&~~$\Delta E $ (MeV)~\\
\hline
45 &1.0000&0.0004&0.9196&1.8718\\
50 &1.0000&0.0004&0.8977&2.6744\\
55 &0.9996&0.0109&0.8759&3.6100\\
60 &0.6067&14.862&0.5127&18.908\\
\hline\hline \end{tabular}
\end{center}
\end{table}

The treatment of the irreversible GDR decay (with a spreading width
of 6 MeV) to the bath of surrounding complex states (represented by
$|Y \rangle$) was included by switching on the coupling of the
intrinsic inelastic state $|2 \rangle$ to $|Y \rangle$. This is the
dynamical scheme of the right hand panel in Figure \ref{Figure0}.
Unlike the coupling to $|X \rangle$, a major part of the inelastic
excitation of the system gives access to $|Y \rangle$  before the
wave packet encounters the fusion barrier. The onset of decoherence,
the purity of the density matrix, and the associated energy
dissipation are shown in the right hand entries in Table \ref{Table}
and, as a function of time evolution, in Figure 2(b), the latter for
$E_0=45$ MeV.

Summing over all internal states of the density matrix gives the
total diagonal elements in coordinate space, which represent the
wave-packet at a given time. Snapshots of the wave-packet in the
interaction region (for $E_0$ = 45 MeV) are shown in Fig. 3. The
curves are shown for times $t$ = 0 (dashed; the initial state), $400
\times 10^{-22}$ s (dotted; near to the time of closest approach)
and $700 \times 10^{-22}$ s (thin solid; post the collision). The Figure
shows the results from (a) the coupling to state $|X \rangle $ , and
(b) to both $|X \rangle $ and $|Y \rangle$. When the wave-packet
tunnels into the pocket (from dashed to dotted lines), the short
range coupling to $|X \rangle $ leads to trapping of flux from $|1
\rangle $ and $|2 \rangle $ inside the potential pocket. This
reveals itself as an unchanging probability for radii $r < 7.5$ fm
as the main body of the wave-packet leaves the interaction region
(from the dotted to the thin solid lines) in Figure 3(a). The additional
effect of turning on the coupling between states $|2 \rangle $ and
$|Y \rangle $ is to trap probability \emph{under} the barrier, as is
shown in Figure 3(b). This reduces the component of the wave-packet
that reaches the potential pocket, inhibiting the quantum tunneling.

\begin{figure}
\begin{center}
\includegraphics[width=0.40\textwidth,angle=0]{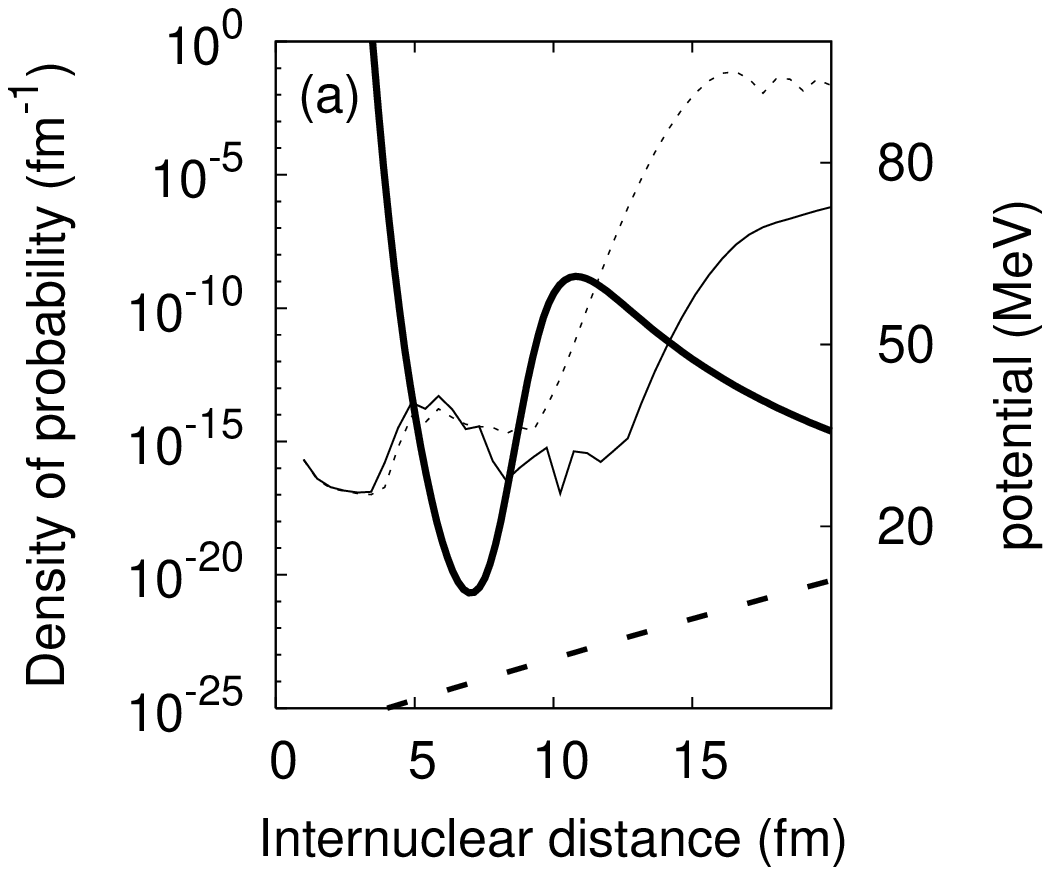}
\includegraphics[width=0.40\textwidth,angle=0]{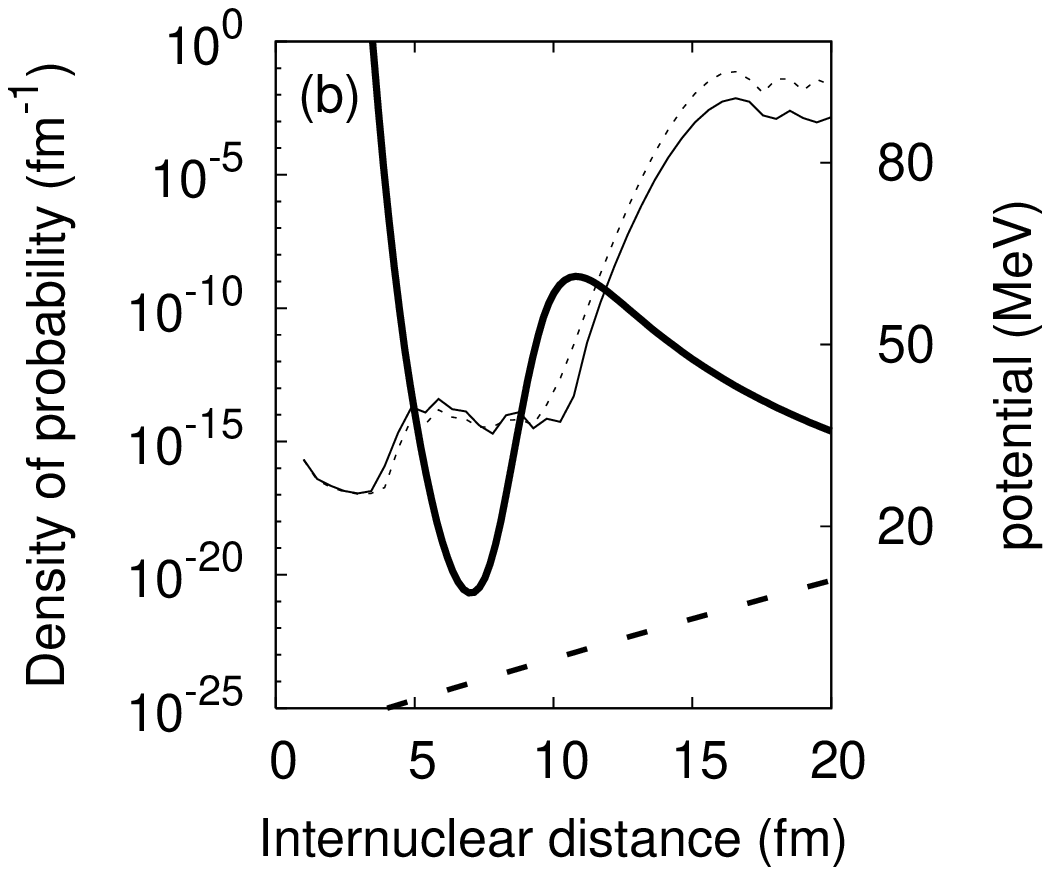}
\caption{The inter-nuclear potential (thick curve) and the time
evolution of the wave-packet for $^{16}$O + $^{144}$Sm at $E_0 = 45$
MeV: (a) including only coupling of auxiliary state $|X \rangle $,
(b) including coupling of the states $|X \rangle$ and $|Y \rangle$.
The wave packet is plotted at times $t=0$ (dashed curve), $400
\times 10^{-22}$ s (dotted curve) and $700 \times 10^{-22}$ s (thin
solid curve).} \label{Figure2}
\end{center}
\end{figure}

\begin{figure}[h]
\begin{center}
\includegraphics[width=0.47\textwidth,angle=0]{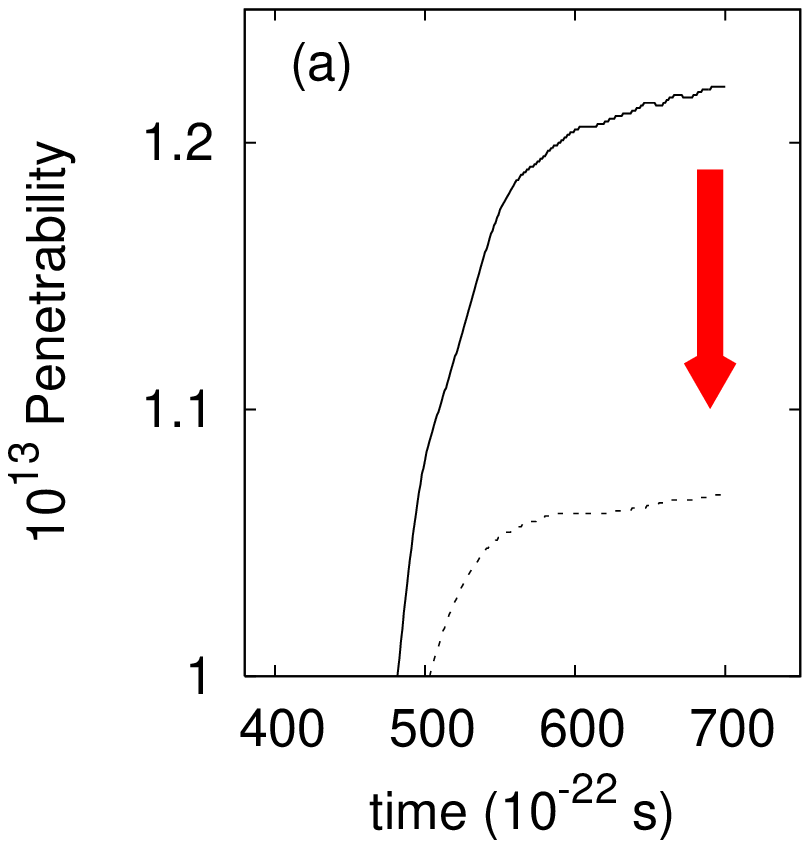}
\includegraphics[width=0.40\textwidth,angle=0]{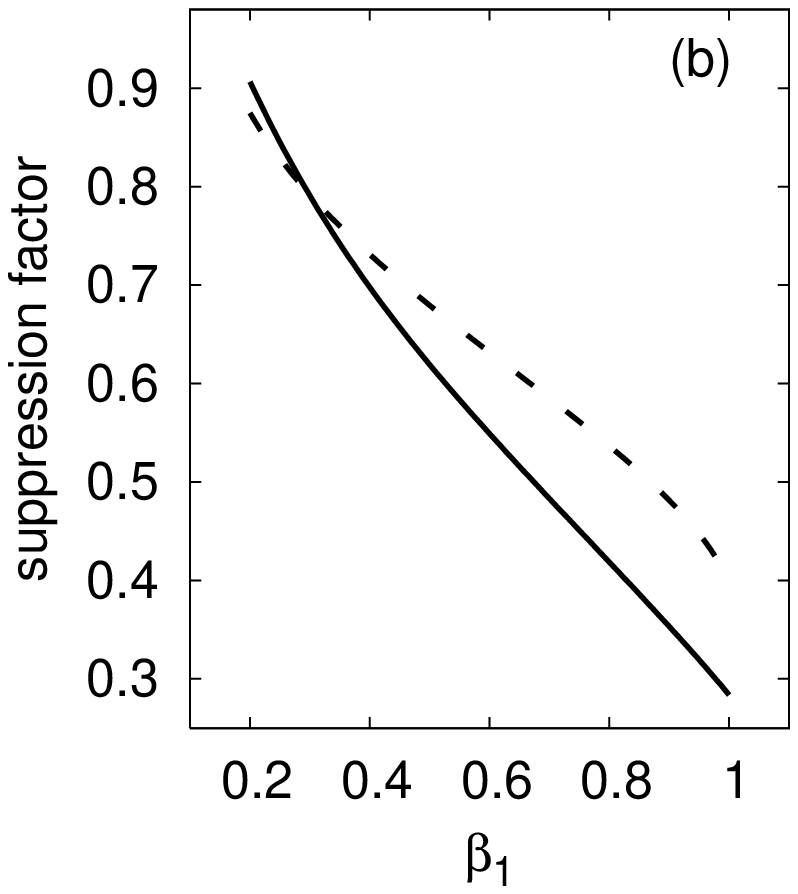}
\caption{(Color online) (a) Time-dependence of the probability
trapped in $|X \rangle$ for $E_0=45$ MeV. The full curve includes
states $|1 \rangle$, $|2 \rangle$ and $|X \rangle$. The dotted curve
adds the irreversible decay of $|2 \rangle $ to $|Y \rangle $. The
calculations are for $\beta_1=0.2$. (b) Calculated suppression of
the probability trapped in $|X \rangle$ as a function of the assumed
$\beta_1$ value for $E_0=45$ MeV (dashed curve) and 55 MeV (solid
curve).} \label{Figure3}
\end{center}
\end{figure}
The probability trapped under the fusion barrier is associated with
GDR collective vibrational energy being irreversibly removed from
the coherent dynamics into innumerable channels (heat). This is then
no longer available for relative motion, or tunneling. Such energy
loss can be correlated with deep inelastic processes, seen
experimentally, that compete with fusion in reactions involving
heavy nuclei \cite{Wolfs}.

Fig.\ \ref{Figure3}(a) shows the time evolution of the probability
trapped in the potential pocket, state $|X \rangle$, for $E_0=45$
MeV. We comment that, when including the inelastic channel $|2
\rangle$ but not $|Y \rangle$, the nucleus-nucleus potential
renormalization leads to the expected enhanced penetrability from
the inelastic channel coupling, compared to the purely elastic ($|1
\rangle$ plus $|X \rangle$) calculation. The decoherent dynamics due
only to environment $|X \rangle$ gives the (full curve). By
comparison, the calculation that also includes the GDR doorway-state
decay to $|Y\rangle$ leads to a suppression (dotted curve and arrow)
of the population of state $|X \rangle$. Additional irreversible
processes other than excitation of the GDR are also likely to
contribute to the deep-inelastic yield, such as complicated
multi-nucleon transfers \cite{Rehm}. To simulate these very simply,
the assumed state $|1 \rangle$ to $|2 \rangle$ coupling strength was 
increased. Fig.\ \ref{Figure3}(b) shows the dependence of the
calculated tunneling suppression on the assumed $\beta_1$ strength
for $E_0=45$ (dashed curve) and 55 MeV (solid curve), where we note
that larger $\beta_1$ result in both an increase in the strength and
the range of the coupling formfactor to the inelastic state $|2
\rangle$.

\section{Summary}

A quantum dynamical model based on time-propagation of a
coupled-channels density matrix has been presented and is shown to
describe the transition from pure state (coherent) to mixed state
(decoherent and dissipative) dynamics during a nuclear collision.
The calculations exhibit both decoherence and energy dissipation and
so go beyond coherent coupled-channels approaches. Decoherence,
originating here from the irreversible decay of a giant-dipole
vibrational state of the heavy target nucleus to surrounding states,
is shown to result in hindrance of quantum tunneling. Developments
of the model calculations to include (a) non-zero relative orbital
angular momenta between the reactants, (b) additional intrinsic
channels, and (c) a more detailed consideration of other processes,
such as multi-nucleon or cluster transfer reactions, are necessary
in order to confront experimental measurements.

\begin{acknowledgments}
We thank J.O. Newton and L.R. Gasques for constructive suggestions.
Support from an ARC Discovery grant and the UK Science and
Technology Facilities Council (STFC) Grant No. EP/D003628/1 is
acknowledged.
\end{acknowledgments}

\end{document}